\documentclass[journal]{IEEEtran}

\IEEEoverridecommandlockouts

\usepackage{cite}
\usepackage[normalem]{ulem}
\usepackage{amsmath,amssymb,amsfonts}
\usepackage{algorithmic}
\usepackage{graphicx}
\usepackage{textcomp}
\usepackage{tabularx}
\usepackage{multirow}
\usepackage{booktabs}
\usepackage{lscape}
\usepackage{longtable}
\usepackage{amsbsy}%
\usepackage{makecell}
\usepackage{color, colortbl}
\usepackage{enumitem}
\usepackage{subfigure}
\usepackage{xcolor}
\usepackage{dirtytalk}
\usepackage{soul}
\usepackage{comment}
\usepackage[ddmmyyyy]{datetime}

\usepackage{tikz}
    \usetikzlibrary{shapes.arrows}
    \usetikzlibrary{plotmarks}
    \usetikzlibrary{patterns}
    \usepackage{tikzscale}

\usepackage{pgfplots}
    \usepgfplotslibrary{fillbetween}
\usepackage{adjustbox}
\usepackage{gincltex}

\usepackage{filecontents}
\usepackage{authblk}
\def\BibTeX{{\rm B\kern-.05em{\sc i\kern-.025em b}\kern-.08em
		T\kern-.1667em\lower.7ex\hbox{E}\kern-.125emX}}

\pgfplotsset{compat=1.14}

\def\BibTeX{{\rm B\kern-.05em{\sc i\kern-.025em b}\kern-.08em
    T\kern-.1667em\lower.7ex\hbox{E}\kern-.125emX}}

\newcolumntype{C}[1]{>{\centering\arraybackslash}p{#1}}
\newcolumntype{L}[1]{>{\arraybackslash}p{#1}}

\definecolor{cobalt}{rgb}{0.0, 0.28, 0.67}
\definecolor{bluegray}{rgb}{0.4, 0.6, 0.8}
\definecolor{babyblueeyes}{rgb}{0.63, 0.79, 0.95}
\definecolor{green}{rgb}{0, 0.5, 0}

\newcommand{\olga}{\textcolor{black}}
\newcommand{\gf}[1]{\textcolor{black}{{#1}}}
\newcommand{\salwa}{\textcolor{black}}

\makeatletter
\pretocmd\@bibitem{\color{black}\csname keycolor#1\endcsname}{}{\fail}
\newcommand\citecolor[1]{\@namedef{keycolor#1}{\color{blue}}}
\makeatother
\usepackage{glossaries}
\newacronym{4g}{4G}{Fourth Generation}
\newacronym{5g}{5G}{Fifth Generation}
\newacronym{6g}{6G}{Sixth Generation}
\newacronym{ntn}{NTN}{non-terrestrial network}
\newacronym{a-ntn}{a-NTN}{aerial NTN}
\newacronym{s-ntn}{s-NTN}{space NTN}
\newacronym{m-ntn}{m-NTN}{maritime non-terrestrial network}
\newacronym{6g-mn}{6G-MN}{6G maritime network}
\newacronym{tn}{TN}{terrestrial network}
\newacronym{iab}{IAB}{integrated access and backhaul}
\newacronym{nr}{NR}{New Radio}
\newacronym{agv}{AGV}{automated guided vehicle}
\newacronym{mmwave}{mmWave}{millimeter-wave}
\newacronym{mec}{MEC}{multi-access edge computing}
\newacronym{ict}{ICT}{information and communications technology}
\newacronym{ml}{ML}{machine learning}
\newacronym{ai}{AI}{artificial intelligence}
\newacronym{lte}{LTE}{Long Term Evolution}
\newacronym{3gpp}{3GPP}{Third Generation Partnership Project}
\newacronym{itu}{ITU}{International Telecommunication Union}
\newacronym{usv}{USV}{unmanned surface vehicle}
\newacronym{imo}{IMO}{International Maritime Organization}
\newacronym{ais}{AIS}{automatic identification system}
\newacronym{iala}{IALA}{International Association of Marine Aids to Navigation and Lighthouse Authorities}
\newacronym{gmdss}{GMDSS}{global maritime distress and safety system}
\newacronym{vhf}{VHF}{very high frequency}
\newacronym{vdes}{VDES}{very high frequency data exchange system}
\newacronym{iot}{IoT}{Internet of Things}
\newacronym{embb}{eMBB}{enhanced mobile broadband}
\newacronym{urllc}{URLLC}{ultra-reliable low-latency communications}
\newacronym{mmtc}{mMTC}{massive machine-type communications}
\newacronym{d2d}{D2D}{device-to-device}
\newacronym{prose}{ProSe}{proximity services}
\newacronym{ue}{UE}{User equipment}
\newacronym{lpwa}{LPWA}{low-power wide-area}
\newacronym{ltem}{LTE-M}{LTE machine-type communications}
\newacronym{nbiot}{NB-IoT}{narrowband IoT}
\newacronym{redcap}{RedCap}{reduced capabilities}
\newacronym{xr}{XR}{extended reality}
\newacronym{ar}{AR}{augmented reality}
\newacronym{mr}{MR}{mixed reality}
\newacronym{vr}{VR}{virtual reality}
\newacronym{npn}{NPN}{Non-public network}
\newacronym{lan}{LAN}{local area network}
\newacronym{qos}{QoS}{quality of service}
\newacronym{rat}{RAT}{radio access technology}
\newacronym{marcom}{MARCOM}{maritime communication}
\newacronym{b5g}{B5G}{Beyond 5G}
\newacronym{vnf}{VNF}{virtual network function}
\newacronym{lstm}{LSTM}{long short-term memory}
\newacronym{dnn}{DNN}{deep neural network}
\newacronym{haps}{HAPs}{high-altitude platforms}
\newacronym{nlos}{NLOS}{non-line-of-sight}
\newacronym{etsi}{ETSI}{European Telecommunications Standards Institute}
\newacronym{mcptt}{MCPTT}{mission-critical push-to-talk}
\newacronym{mcdata}{MCData}{mission-critical data}
\newacronym{mcvideo}{MCVideo}{mission-critical video}
\newacronym{dbn}{DBN}{deep belief network}
\newacronym{fl}{FL}{federated learning}
\newacronym{lp}{LP}{linear programming}
\newacronym{dl}{DL}{Deep learning}
\newacronym{sar}{SAR}{search and rescue}
\newacronym{ospf}{OSPF}{open shortest path first}

\usepackage{textcomp}
\newcommand\copyrighttext{%
  \footnotesize This article has been accepted for publication in a future issue of IEEE Network, but has not been fully edited. Content may change prior to final publication. \\
  \textcopyright 2022 IEEE. Personal use of this material is permitted.  Permission from IEEE must be obtained for all other uses, in any current or future media, including reprinting/republishing this material for advertising or promotional purposes, creating new collective works, for resale or redistribution to servers or lists, or reuse of any copyrighted component of this work in other works.
  }
\newcommand\copyrightnotice{%
\begin{tikzpicture}[remember picture,overlay]
\node[anchor=north,yshift=-5pt,] at (current page.north) {\fbox{\parbox{\dimexpr\textwidth-\fboxsep-\fboxrule\relax}{\copyrighttext}}};
\end{tikzpicture}%
}

\begin{document}

\title{AI-Aided Integrated Terrestrial and Non-Terrestrial 6G Solutions for Sustainable Maritime Networking}

\author{Salwa Saafi, 
        Olga Vikhrova, %~\IEEEmembership{Member,~IEEE}, 
        G\'{a}bor Fodor, %~\IEEEmembership{Member,~IEEE},
        Jiri Hosek, %~\IEEEmembership{Member,~IEEE},
        and Sergey Andreev %~\IEEEmembership{Member,~IEEE}
\thanks{Salwa Saafi (corresponding author, email: saafi@vut.cz) is with Brno University of Technology and Tampere University.
Olga Vikhrova, and Sergey Andreev are with %the Faculty of Information Technology and Communication Sciences, 
Tampere University, Finland. G\'{a}bor Fodor is with Ericsson Research and KTH Royal Institute of Technology. Jiri Hosek is with the Department of Telecommunications, Brno University of Technology, Czech Republic.}
}

\maketitle

\copyrightnotice

\begin{abstract}
The maritime industry is experiencing a technological revolution that affects shipbuilding, operation of both seagoing and inland vessels, cargo management, and working practices in harbors. This ongoing transformation is driven by the ambition to make the ecosystem more sustainable and cost-efficient. Digitalization and automation help achieve these goals by transforming shipping and cruising into a much more cost- and energy-efficient, and decarbonized industry segment. The key enablers in these processes are always-available connectivity and content delivery services, which can not only aid shipping companies in improving their operational efficiency and reducing carbon emissions but also contribute to enhanced crew welfare and passenger experience. Due to recent advancements in integrating high-capacity and ultra-reliable terrestrial and non-terrestrial networking technologies, ubiquitous maritime connectivity is becoming a reality. To cope with the increased complexity of managing these integrated systems, this article advocates the use of artificial intelligence and machine learning-based %data-driven 
approaches to meet the service requirements and energy efficiency targets in various maritime communications scenarios.
%\gf{\it GF: The title refers to "sustainable", so we need to mention sustainability also in the abstract.}
\end{abstract}

%\begin{IEEEkeywords}
%6G, AI, maritime, sustainability, energy efficiency, (non-)terrestrial networks.
%\end{IEEEkeywords}

\vspace{-0.6cm}

\section{Introduction}\label{sec1}
The maritime industry anticipates a substantial increase in the number of operating vessels, 
% and the creation of 
new harbors, and routes worldwide in response to the trade facilitation initiatives supported by the World Trade Organization. 
\gf{These initiatives aim} to speed up international trade and unlock inclusive economic development. As a fast-growing sector, this industry is also experiencing external pressures \olga{due to} environmental concerns. Today's greenhouse gas emissions from shipping are estimated as 2.6\% of \gf{the} total global emissions, which is the equivalent of emissions from a large country~\cite{itf2018decarbonising}. To cope with this escalation, environmental sustainability and digital inclusion practices become fundamental and have to be incorporated across all maritime operations.

For \textit{sustainable industry}, the maritime sector needs to go through an extensive optimization and evolution toward
fully autonomous, globally connected, and digitalized operations with zero-emissions~\cite{ecmar}.
The success of automation of the maritime industry relies heavily on \textit{dynamic networking} and \textit{\gls{ai}} technologies that foster the emerging applications, such as intelligent harbors, remote on-board maintenance, and autonomous docking.
These require unprecedentedly high data rates, large-scale connectivity between a large number of dissimilar terminals, life-long learning and inference at the smart end-points, and seamless operation of terrestrial and non-terrestrial networks.

Recent findings showed that nearly 90\% of data generated on-board never leaves the deck, which means that operators are missing out on valuable insights and analytics for improved logistics, cost of maintenance, and resource utilization~\cite{vodafone2019}. %\gf{\it GF: a reference here would be nice.}
Owing to recent advances in satellite technology, the number of connected vessels has doubled over the last 5 years, but only 75\% of vessels have on-board Internet access today. Satellite-based backhauling remains extremely costly and inherently limited, thus presenting serious challenges for the widespread technology adoption to meet the growing needs of the maritime industry. %Vessels spend about 60\% of time within reach of a terrestrial network while in port or in coastal waters. 
However, existing maritime communication systems offer dedicated %-- typically narrowband -- 
services opportunistically (e.g., in proximity to coastal infrastructure), rather than genuinely \gf{inter-connecting} humans, vessels, and ports into a holistic ecosystem.

While attempting to integrate several non-terrestrial networks into a unified infrastructure to facilitate the demanding intelligent broadband services for maritime operations, cellular \gls{5g} systems become increasingly convoluted and energy consuming, which risks compromising the fundamental need for sustainability~\cite{gradiant2019}.
By contrast, \gls{6g} technologies are envisaged not only as those employing higher frequencies (e.g., \gls{mmwave} and Terahertz bands) to achieve extreme throughputs, 
%more stringent service requirements, 
but also as solutions capable of supporting \gls{ai}-aided closed-loop automation. %, which may be optimized for data-driven operation. 
Such systems are expected to enable the ultimate potential of the zero-touch architecture proposed by the \gls{etsi} for fully automated networks~\cite{etsizero}.

%\rev{In this work, we characterize the \textit{\gls{6g-mn}} as a prospective 
%\gls{6g} use case requiring careful design under the surge in communication 
%and computation demands fueled by the \gls{ai} and \gls{6g} convergence.
%\sout{Our approach to achieving beyond additive growth in energy efficiency is a holistic pursuit of (i) distributed system intelligence where \gls{ml} and inference are employed across the device-edge-cloud continuum, and (ii) dynamic networking where cells and links are created on-demand employing high-rate communications capabilities.} 
%To achieve beyond additive growth in energy efficiency, one needs a holistic pursuit of 
In this article, we first examine the most important of the emerging maritime use cases, which call for a careful design of a \textit{\gls{6g-mn}} that facilitates communication and computation applications.
The key components of \gls{6g-mn}s are dynamic networking -- where communication links are created and activated on-demand -- and distributed intelligence, where learning and inference are employed at different levels of the system. In our supportive study, we demonstrate how \gls{ml} can aid and outperform traditional model-based approaches for energy-efficient topology management and scheduling in dynamic maritime networks. We then identify communication- and learning-related challenges in future \gls{ai}-aided \gls{6g-mn}s.

The rest of this article is organized as follows.
We first introduce the rationale for building future maritime communication systems around cellular networks and discuss the essential maritime use cases. We then review the challenges of sustainable operation in \gls{6g-mn}s and highlight the key role of \gls{ai} for network-wide optimization in terms of topology management and resource allocation. %We illustrate the benefits of employing \gls{ml} methods with numerical examples.
We also summarize the open issues related to the integration of \gls{ai}-based solutions in \gls{6g-mn}s and offer concluding remarks and future perspectives on this work in the final section.

\section{Cellular-Enabled Maritime Networks}\label{sec2}

\subsection{Existing Maritime Communication Systems}\label{sec2.1}

Targeting navigation safety and maritime environment protection,
the \gls{imo} and the \gls{itu} cooperatively launched the \gls{gmdss}.
The latter includes a set of terrestrial and satellite radio technologies employed
%in safety procedures such as the rescue of persons and vessels in distress \cite{itu2020modern}.
for people and vessel rescue in distress.
To further improve ship-to-ship and ship-to-shore navigation accuracy,
\gf{the} \gls{imo} introduced the \gls{ais}, which complements marine radars with tracking information for vessel collision avoidance and better situational awareness~\cite{itu2020modern}.

Despite being an effective technology for %the support of assistance to
navigation assistance and %response to
maritime emergency services, the \gls{ais} provides \salwa{low data rate communications for the exchange of basic navigation parameters such as speed, position, and direction~\cite{gradiant2019}.}
%solely voice and low data rate communications \cite{gradiant2019}.
This limitation motivated the \gls{iala} to develop their own \gls{vdes}.
Building upon \gls{ais} capabilities, \gls{vdes} encompasses several communication subsystems
aiming to provide higher data rates, enhance the operating ranges by the integration of satellite components,
augment security mechanisms,
and support new maritime use cases such as e-Navigation \cite{gradiant2019}.
The concept was introduced by the \gls{imo} to %characterize the harmonization of
harmonize maritime navigation systems in %the support of
offshore and coastal regions~\cite{itu2020modern}.

Although regulatory bodies continue to improve the systems discussed above, the ongoing evolution toward \gf{a digitalized} maritime industry poses new challenges to maritime communication system design. With the increasing number of vessels, growing level of ship autonomy, and widespread adoption of \gls{iot} technologies, novel connectivity solutions need to ensure cost-efficiency, scalability, and service availability~\cite{iala2017mrcp}.
%To offer different services, maritime networks will integrate heterogeneous communication systems, as the \gls{iala} identified in its maritime radio communication plan.
%\gf{\it GF: I am not sure I really understood the previous sentence. I think what we would like to say is that there is a need for multiple connectivity solutions.}
% available for the support of maritime networks.
Such communication systems for modern maritime operations have to support not only the existing e-Navigation and \gls{gmdss} services, but also the emerging broadband and low-latency applications discussed in the sequel.

\subsection{Emerging 6G-MN Use Cases}\label{sec2.2}

\begin{figure}[bthp]
\centering
\includegraphics[width=\columnwidth]{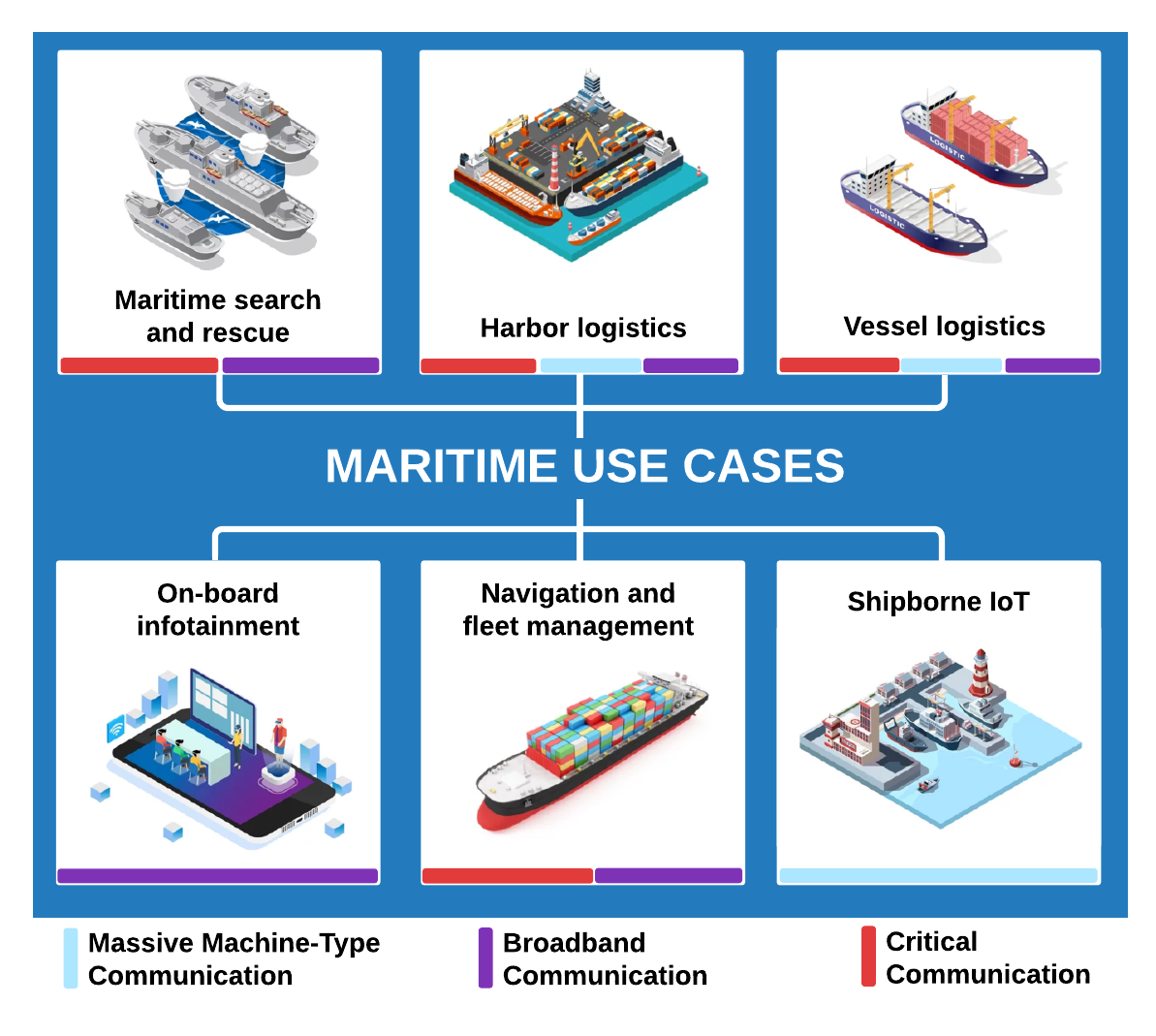}
\caption{Use case categories in 6G-MNs}
\label{fig:usecases}
\end{figure}

Based on the \gls{imo} and \gls{ais} specifications and our understanding of market trends, the emerging maritime use cases can be grouped into six categories as illustrated in Fig.~\ref{fig:usecases}. This figure also maps the use cases onto connectivity requirements in terms of \gls{5g} service classes, namely, \gls{embb}, \gls{urllc}, and \gls{mmtc}. 

The navigation and fleet management applications facilitate the exchange of telematic information 
for enhanced situational awareness and maritime fleet management. 
They provide mission-critical services for vessels of different types (i.e., cargo, law-enforcement, research, commercial, and leisure) 
and shore-based traffic management organizations %, e.g., vessel traffic service centers,
using high-speed broadband links.

Shipborne \gls{iot} aims to improve on-board operation and navigation by communicating vessel's speed, fuel consumption, and carbon dioxide emission information to the on-board sensor fusion systems. Narrowband and massively deployed sensors generate abundant machine-type raw data for subsequent analysis and feature extraction. 
Vessel logistics is another shipborne use case category where the crew utilizes on-board communications for staff coordination and supply management and in the case of internal emergencies. 
On top of that, on-board infotainment provides passengers with access to video streaming, gaming, and interactive applications. 
These new shipborne use case categories comprise the requirements of broadband, critical, and massive machine-type communications, thus making the system highly heterogeneous in terms of traffic patterns, \gls{qos} requirements, and device capabilities.

As the name implies, the use case category of maritime \gls{sar} provides medical emergency and ``Man Overboard'' rescue services that entail broadband and mission-critical applications to connect users in distress, on vessels, and around shore-based facilities.
To continue with shore-based applications, harbor logistics %is a use case category that
offers a range of services for planning, organization, and inspection of harbors and industrial port operations. 
Ship loading/unloading coordination, asset tracking, warehouse management, short-sea, and feeder shipping 
can be monitored via these services. Similarly to vessel logistics use cases, harbor logistics %can have the requirements of
may involve all three types of communication regimes.

\begin{table*}[ht]
\centering
\caption{Overview of enabling solutions for maritime use cases}
\label{tab1}
%\begin{tabular}{|p{1.5cm}|p{2.5cm}|p{3cm}|} \hline
\begin{tabular}{|p{2.2cm}|p{0.3\textwidth}|p{0.5\textwidth}|} \hline
\textbf{Solution}  & \textbf{Related concepts} & \textbf{Relevant \gls{6g-mn} use cases} \\ \hline \hline
5G service classes &  eMBB, URLLC, mMTC & Broadband, critical, and massive machine-type communications for all \gls{6g-mn} use cases \\  \hline
NR NTN &  Satellite communication networks, unmanned aerial systems, HAPs &  Coverage extension, network access in offshore areas and NLOS scenarios \\ \hline
Mobile IAB &  Wireless backhaul, IAB-donor, IAB-nodes &  Capacity improvement within a vessel, coverage extension in offshore areas \\ \hline
MCX services &  MCPTT, MCData, MCVideo, off-network MCX & Rescue services, response to shipborne emergencies \\ \hline
LTE/NR sidelink & D2D communications, ProSe, UE-to-network relay  &  Ship-to-ship communications for navigation and collision avoidance, sensor group communications for shipborne IoT, relaying for coverage extension \\ \hline
MEC &  Edge cloud servers, computation task offloading & Low latency and low energy consumption in shore-based and offshore applications \\ \hline
Cellular LPWA & LTE-M, NB-IoT, mMTC & Shipborne cellular IoT \\ \hline
NR RedCap  &  Industrial sensors, surveillance cameras, wearables & Using data collected by RedCap devices in several \gls{6g-mn} use cases \\ \hline
5G XR   &   AR, MR, VR &  XR for mission-critical maritime \gls{sar}, XR conferencing for vessel logistics \\ \hline
5G LAN &  5G LAN-type access, enterprise network communications & On-board services including vessel logistics and infotainment \\ \hline
NPN   &   Public network-integrated NPNs, standalone NPNs & Smart ports/harbors, enhanced on-board connectivity \\ \hline
Positioning &  High-accuracy positioning, RAT-dependent, RAT-independent, hybrid solutions & Vessel location awareness for navigation and fleet management, indoor positioning services for staff management \\ \hline
Platooning & Cooperative platoons, autonomous vessels & Short-sea shipping and feeder services within harbor logistics use cases \\ \hline
\end{tabular}
\end{table*}

\subsection{Prospective Solutions for 6G-MNs}\label{sec2.3}

Several attempts to interconnect vessels in coastal waters and build a bridge to the port have been successful by virtue of cellular coverage. However, due to the limited capacity of communication links, existing systems for maritime communications provided by, for example, Cellnex Telecom or Telenor Maritime, fail to cover deep offshore areas and support delay-critical and bandwidth-hungry use cases. By contrast, \gls{5g} and beyond networks can provide a flexible and adaptive mobile communication platform for the modernization 
and long-term support of the maritime industry not only in coastal but also in offshore areas, as confirmed by the \gls{3gpp} in TR 22.819.

Initial \gls{6g} systems are to be primarily supported by the existing \gls{5g} infrastructures, thus benefiting from the advancement of cellular technologies that can foster the deployment of even more agile and intelligent applications discussed above.
%the software-defined network, virtual network function, and network slicing technologies.
%However, compared with \gls{5g} networks, \gls{6g} networks require to support new services with the very stringent requirements in terms of data rates, latency, reliability, connectivity, and, more importantly, energy-efficiency, scalability, and network agility.
Enabling solutions, related cellular concepts, and relevant \gls{6g-mn} use cases are offered in Table~\ref{tab1}. %These solutions can foster the deployment of even more agile and intelligent applications discussed above in \gls{6g} ecosystem.

Hybrid satellite--terrestrial networks have been utilized to complement the high capacity of shore-based systems by the wide-area coverage of satellite communications. In addition to satellite access, \gls{nr} technologies can support other non-terrestrial access components, such as \gls{haps} as identified by \gls{3gpp} in TR 38.811. By employing solutions for \gls{nr} to support \gls{ntn}, 
\gls{6g-mn}s can benefit from the 5-layer architecture for \gls{6g} setups as proposed in \cite{yanikomeroglu2018integrated} 
to extend the coverage of terrestrial systems and provide access to maritime services in offshore areas and \gls{nlos} scenarios.

In addition to \gls{nr} \gls{ntn}, an attractive technology for coverage and capacity extension in terrestrial \gls{5g} and beyond networks is \gls{iab}, which was introduced by \gls{3gpp} TR 38.874 in Release 16. It is based on using part of the wireless access spectrum for the backhaul connections between remote base stations. While fixed \gls{iab}-nodes can extend the capacity of coastal \gls{6g-mn}s, mobile \gls{iab}, as suggested among the \gls{3gpp} Release 17 enhancements, can be employed for dynamic backhaul solutions, thus shaping mobile maritime mesh networks for both coverage and capacity extension. This can be considered as a more flexible, scalable, and affordable broadband access option for a wide range of cloud services.

Further, \gls{3gpp} public safety services, including \gls{mcptt}, \gls{mcdata}, and \gls{mcvideo}, can be applied for emergency services. These services can also be provided \salwa{using the off-network mode in areas} wherein cellular coverage is temporarily unavailable or network performance is limited in terms of capacity or latency.
%These services can also be provided in the out-of-band mode for ship-to-ship and group on-board communications over deep offshore areas wherein cellular coverage is temporarily unavailable or network performance is limited in terms of capacity, latency, or user throughput.

\gls{nr} sidelink for \gls{d2d} communications and \gls{prose} can help avoid collisions as part of the navigation and fleet management use case. \salwa{Sidelink-based ship-to-ship communications allow vessels to assist maritime coordination centers during \gls{sar} missions.} \gls{ue}-to-network relay, as another form of \gls{d2d} wherein an indirect network connection is provided by a relay \gls{ue}, can extend the coverage of terrestrial systems for shipborne applications in coastal waters.

In \gls{5g} networks, computation resources are moved closer to the end devices. Heavy computation tasks and raw data can be offloaded to proximate \gls{mec} servers, thus minimizing end-to-end transmission delays and energy consumption. Cellular \gls{lpwa} technologies, namely, \gls{ltem} and \gls{nbiot}, can be deployed in \gls{6g-mn}s 
to support the use cases with \gls{mmtc} requirements. In a similar context, the need for \gls{nr} devices with \gls{redcap}, such as on-board industrial sensors, video surveillance cameras, and wearables, has been addressed by \gls{3gpp} in TR 38.875. Future \gls{6g-mn}s can benefit from the \gls{nr} \gls{redcap}-enabled services for enhanced vessel navigation and monitoring, harbor logistics, and remote on-board assistance. The latter example refers to one of the \gls{xr} use cases defined in TR 26.928, which includes augmented reality-guided assistance in remote locations, mixed reality-based sharing, and virtual reality-based telepresence collaboration.

\gls{npn} and \gls{lan}-type access for industrial \gls{iot} use cases has already been considered by \gls{3gpp} for maritime scenarios in TR 22.819. With the planned enhancements in Release 17, \gls{npn}s can be employed in smart harbors to build private networks with adequate \gls{qos} and security guarantees. A range of positioning schemes, including \gls{rat}-dependent and \gls{rat}-independent solutions, are ratified in \gls{3gpp} Release 16
specifications and can be used separately or in a hybrid manner to meet the required positioning accuracy in \gls{6g-mn} use cases. Accurate positioning can also enable future maritime platooning services. Aiming to improve navigation safety and reduce fuel consumption, autonomous vessels can move in cooperative platoons and be employed in feeder services to manage short-sea shipping from hub ports to feeder ports in inland waterways~\cite{thormann2020safe}.

Recently, recognizing the high interest from maritime stakeholders, \gls{3gpp} has officially included the work on \gls{marcom} services over cellular systems in its beyond Release 16 standardization efforts in TS 22.119. An important challenge that needs to be considered in these systems is energy efficiency as part of the sustainability goal in maritime operations. Even though \gls{5g} \gls{nr} system design offers better bit-per-Joule energy efficiency as compared to the previous generations of mobile technology, a typical \gls{5g} site has nearly 70\% higher energy consumption than a base station deploying a mix of 2G, 3G, and 4G radios due to the use of additional power-hungry components~\cite{shurdi20215g}. Further cell densification together with link heterogeneity can make it difficult to optimize \gls{5g} and beyond deployments in real time, which calls for a more adaptive and less energy consuming system design discussed in the following section.

\begin{figure*}[tbh]
    \centering
    \includegraphics[width=\linewidth]{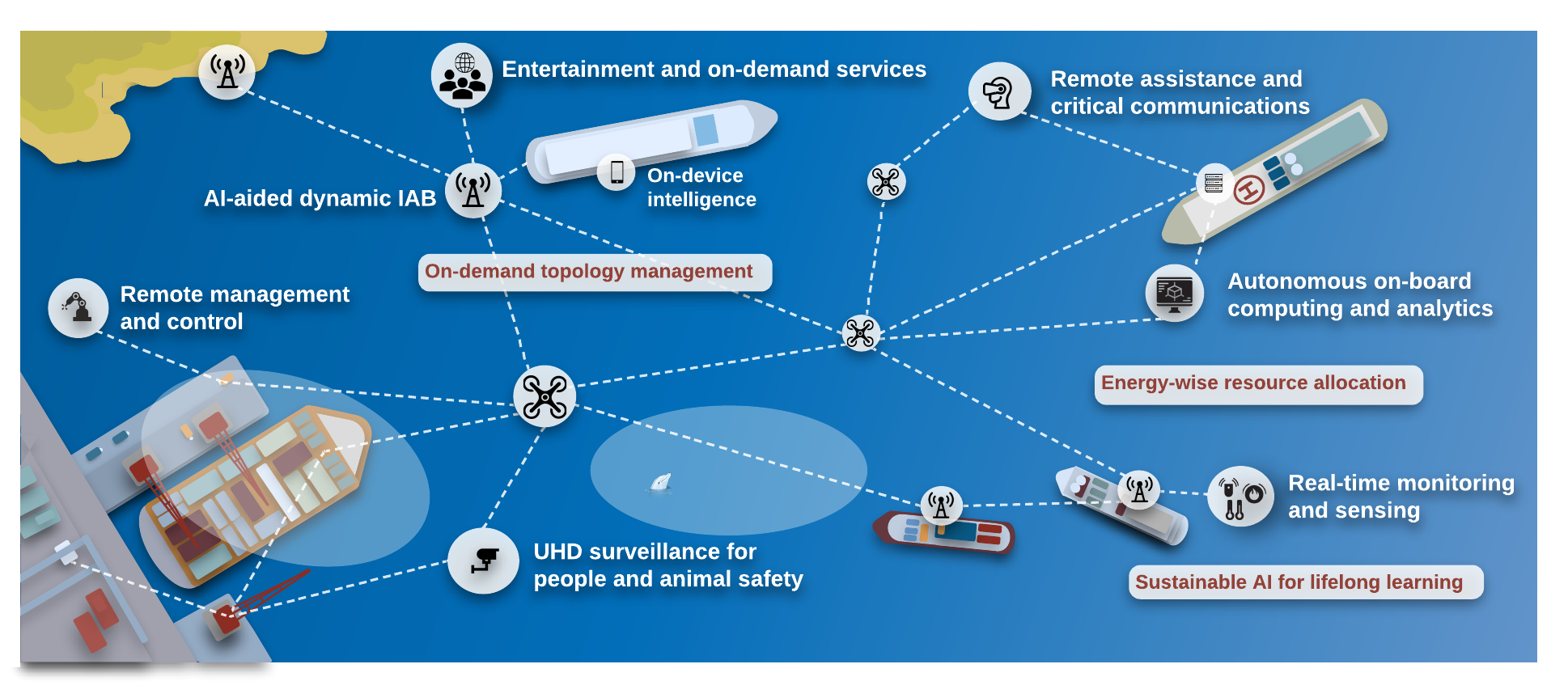}
    \caption{Our vision of a unified, scalable, reliable, and intelligent \gls{6g} system with integrated features for sustainable maritime operations}
    \label{fig:vision}
\end{figure*}

\section{AI for Sustainable Maritime Networking}\label{sec3}

\subsection{Network Sustainability Challenges}\label{sec3.1}

The use cases identified in Fig.~\ref{fig:usecases} are different from the scenarios behind the operation of well-established mobile and vehicular ad-hoc networks. For instance, they can create more stable multi-hop network formations over longer time spans, may need to transmit heterogeneous data over larger communication ranges, and might also include more complex and advanced on-board and in-port communication scenarios similar to industrial \gls{iot} applications.

A large volume of data generated on-board the vessels in deep offshore areas needs to be offloaded to the mainland for efficient port operation, or distributed further within the network to facilitate optimal maritime navigation and logistics. %For instance, sharing detailed weather observations, as well as tidal data or pollution levels, with port authorities helps to be prepared against imminent weather phenomena and can provide valuable input to prevent loss of communication links.
Real-time data collected along the navigation routes are crucial for autonomous %commercial and private 
shipping, as ports and vessels can access this information upon request. This can mitigate deviations from the optimal port operation and failures of communication links. In contrast, infotainment and extended reality-based applications for passengers and crew members of small vessels and large cruisers can require rapid dissemination of heavy content from the land to numerous destinations in different parts of the world.
As these examples suggest, maritime operation targets robust, scalable, high-performance, and adaptive mechanisms for the orchestration of dissimilar services over dynamically formed mobile networks.

In \gls{6g-mn}s, multi-hop communications and flexible mesh topologies can be supported by \gls{iab} and \gls{d2d} solutions. These technologies help overcome the vulnerabilities of highly directional and fast fading links, 
%to inconsistent 3D coverage and abrupt (self-)blockage due to the node mobility or environment changes. They also help 
tolerate increased interference levels, and utilize radio resources more efficiently.
%However, using these solutions leads to a unique challenge of energy-efficient formation and management of dynamic topology in integrated maritime networks.
However, due to node mobility within the unique and challenging integrated maritime infrastructure, \gls{6g-mn}s need to continuously adapt their resource allocation and scheduling policies over dynamic topologies.

Our envisaged \gls{6g-mn} illustrated in Fig.~\ref{fig:vision} comprises a high-performance terrestrial network segment that inherently supports operation and maintenance of smart harbors and industrial ports. The non-terrestrial part encompasses a dynamic \gls{iab} infrastructure, which ensures connectivity bridges between vessels and the terrestrial segment. The former facilitates inland shipping and ubiquitous monitoring of near-shore areas for improved human and marine life safety. An important component of this system is the powerful edge and cloud infrastructure for centralized and distributed learning to enable proactive and resource-wise on-demand operation.

Traditional model-based approaches and optimization algorithms may not be sufficient for satisfying the requirements of the above system. They typically struggle with a lack of timely global information about the system (e.g., channel and buffer states, user mobility, or demand level) to provide optimal control instructions and can thus yield impractical computations due to excessive model dimensions. In turn, data-driven methods are known to efficiently deal with both model and algorithm deficits, and with learning functional relations between different system parameters that are difficult to model. These parameters include (i) user profile data such as device position, mobility, transmission, and energy consumption patterns, (ii) network configuration data encompassing instantaneous link capacity and resource utilization, and node capabilities, and (iii)
service data covering quality of experience, subscription to cooperative learning of a \gls{ml} model, and capabilities for execution of offloaded tasks.

Allowing to extract valuable information from these massive data, \gls{ai} techniques can be used to predict traffic peaks and system demands, detect anomalies or near-overload conditions, and identify nodes or clusters with high energy and resource consumption. This knowledge can alleviate the topology management and scheduling problems and underpin the solution design for self-optimized and automated \gls{6g-mn}s with energy-oriented optimization goals. In what follows, we provide examples of \gls{ai} applications for a more energy-efficient maritime system operation.

\subsection{Energy-Centric Topology Management}\label{sec3.2}

Since both system topology and network load may evolve over time, storage, compute, and spectrum resource allocations have to be provided on-demand and thus periodically re-optimized.
%The most natural way to avoid a prohibitive growth and energy consumption of \gls{6g-mn}s is to provide the necessary resources on-demand. 
To manage on-demand topologies, joint link activation and resource allocation problems need to be solved repeatedly, and potentially compared with previous configurations. As traditional model-based methods can be resource and time consuming, a more agile approach is required to facilitate the repetitive optimization problems. %and, thus, to manage on-demand topologies efficiently consuming less energy.

As an illustrative example, we consider a low-dynamic multi-hop wireless network deployed over 100 km$^2$, where access nodes are uniformly distributed across the given area with the density of $0.5$ nodes per km$^2$. Our goal is to minimize the network energy consumption while delivering heterogeneous traffic originating from cruisers or cargo vessels and rerouted to different destinations (e.g., harbors or other vessels). Therefore, we are aiming at the optimal allocation of time, spectrum, and power along the optimal routing paths. %and optimize power allocation at the \gls{iab} nodes to mitigate interference.
%\gls{ml}-based data-driven approaches facilitate solving the challenging problem of joint power allocation and scheduling channels between \gls{iab} nodes in \gls{6g-mn}s in the presence of multiple heterogeneous traffic flows.

Given that the scale of the above optimization problem increases with the number of nodes and potential routes, the performance of traditional optimization approaches can degrade dramatically. For solving the joint power allocation and link scheduling problem in this system to minimize its overall energy consumption, one needs to know all the possible allocation patterns to employ \gls{lp}. The number of patterns primarily depends on the number of potential links, though not all of them may eventually be used in the optimal solution~\cite{liu2020deep}. 

Since the relation between the network flows and the set of active links in the optimal configuration cannot be obtained analytically, it may be learned by a data-driven method using information about only some of the optimal configurations (e.g., by solving optimization problem analytically for a number of system setups with fewer flows). Once the model is trained, it can efficiently predict which links are critical for the optimal configuration under any new traffic flows in the system. Therefore, by using a \gls{dnn}, particularly its sub-category \gls{dbn}, we reduce the number of links involved in the routing and scheduling decision in a given multi-hop layout, and thereby alleviate the complexity and execution time of finding the optimal configuration.
%\gf{\it GF: Can we talk about the link activation as a random variable ? Or is it the duration of the link activation what is the random variable whose distribution we refer to ?} 

\begin{figure}[bthp]
    \centering
    \includegraphics[width=\columnwidth]{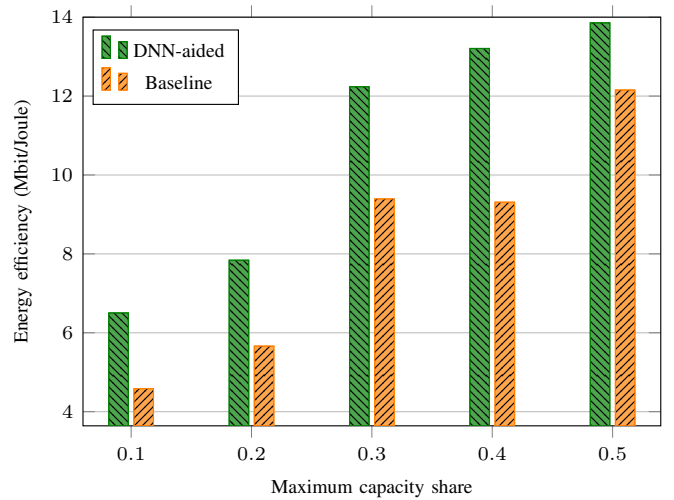}
    \caption{Energy-efficient resource management for \gls{iab}-aided backhaul solution}
    \label{fig:multihop}
\end{figure}

Hence, Fig.~\ref{fig:multihop} demonstrates our system-level simulation results for the energy efficiency of a maritime communication system operating at 3.5 GHz with different traffic loads defined as the maximum system capacity share. These results have been first obtained by using the \gls{dnn}-aided \gls{lp} optimization framework described in~\cite{liu2020deep}, and then compared to the baseline system operation.
%due to the faster and less computational resource consuming re-optimization.

Along these lines, other promising techniques such as \gls{lstm} and auto-encoder can be applied to discover the temporal and spatial correlations in traffic and support the formulation of multi-objective optimization problems. For critical services with ultra-low latency requirements, the challenge of routing is central. As shown in~\cite{kato2017deep}, a supervised \gls{dnn}-aided traffic routing scheme causes much lower overheads than the traditional options such as \gls{ospf}, while guaranteeing acceptable delay.

\subsection{Energy-Efficient Scheduling}\label{sec3.3}

%\salwa{Extremely high frequency} radio suffers from inconsistent 3D coverage and abrupt (self-)blockage situations due to large- and small-scale node mobility. 
Due to large- and small-scale node mobility, as well as interference fluctuations, the reported channel quality may become outdated, misleading, or even lost at the network side. Systematic inaccurate or imperfect knowledge of the channel state may cause significant \gls{qos} and energy efficiency degradation. %, resource wastage, and increases energy consumption on device and network sides. 
Hence, mechanisms for channel quality prediction can help combat different types of link blockage effects and control radio interference while supporting redundant on-demand topologies.

\gls{dl} methods can improve the accuracy of channel reporting and, consequently, avoid inefficient resource utilization in the core network and radio access parts. They are well-suited for capturing non-linear and dynamic relationships between the model input and output data. They also have powerful prediction, inference, and data analysis capabilities owing to the large amounts of data generated by the environment and by the users. In particular, \gls{lstm} can handle time series problems, which makes them attractive for channel quality prediction and capable of alleviating the physical layer imperfections~\cite{yin2020predicting}.

\begin{figure}[bthp]
    \centering
    \includegraphics[width=\columnwidth]{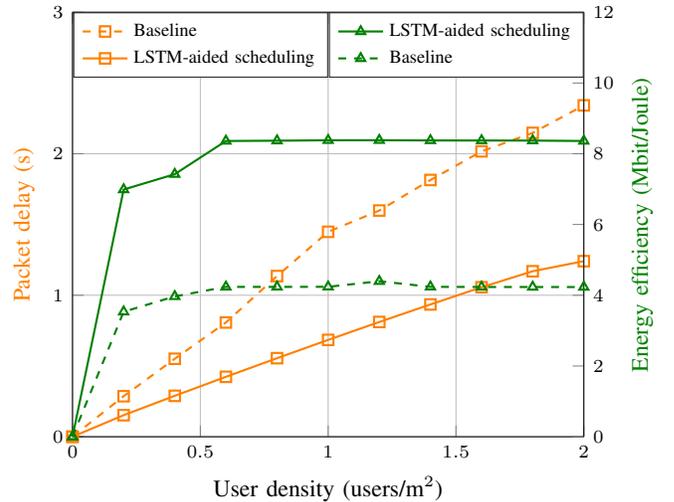}
    \caption{Use of LSTM for energy-efficient real-time scheduling}
    \label{fig:cqi_prediction}
\end{figure}

The results of our system-level simulations summarized in Fig.~\ref{fig:cqi_prediction} demonstrate a significant improvement in energy efficiency by applying \gls{lstm} for resource scheduling. We employ an open-source interface between network simulator-3 and Python-based \gls{ai} frameworks. The latter train the \gls{lstm} model using data generated by the simulator and then return the data from the trained model back to the simulator for testing. Communications over \gls{mmwave} channels in a single-cell network topology with mobile \gls{ue}s ($10 m/s$) are assumed.
%, where training and testing data are exchanged between the two modules.
%The user density is in order of $10^6$ users per km$^2$, while communication over \gls{mmwave} channels is assumed. 
Unlike in the baseline scenario (i.e., without \gls{ml}), in the \gls{lstm}-aided case the base station utilizes the predicted channel quality information when making decisions about scheduling and radio resource allocation~\cite{yin2020predicting}. Not only the system energy efficiency can be improved with better channel quality predictions, but also the end-to-end packet delay may be significantly decreased.

\section{Open Issues in Employing AI for 6G-MNs}\label{sec4}

\subsection{From Centralized to Distributed Learning}\label{sec4.1}

%Sustainability goal of \gls{6g-mn} adds energy-efficiency requirements on top of the recently emerged 3C optimization framework referring to the joint optimization of computation, communication, and control of resources. % in systems empowered with edge intelligence.
%The above optimization targets are new to the cellular community, which has only recently started working on energy-efficient maritime system design~\cite{zeng2020energy}. 
As the adoption of \gls{ai} technologies accelerates, the integration of various monitoring and control systems within a centralized cloud can limit the scalability in such systems. %Further capacity expansion in such systems may also be challenging.
Hence, today’s predominantly cloud-centric \gls{ai} solutions that rely on training and inference in the remote cloud have to be complemented by more energy-efficient, partially distributed, and ultimately fully distributed learning mechanisms where numerous devices collaboratively train a part of a global model~\cite{hosseinalipour2020from}.

%An additional challenge in \gls{ai}-aided \gls{6g-mn}s will be the manner of handling the massive data described in Section~\ref{sec3.1}. Regarding this issue, device-level solutions can be used to avoid the need for transmitting huge amounts of collected data from \gls{ue}s to processing nodes.

%\gf{which calls for a unified and all-in \gls{6g} cellular solution, which can be more appealing to the maritime operators.} \gf{\it GF: I am not sure if these expressions "unified" and "all-in" are clear to the reader at this point. So I think we may need to make a forward reference, "...as it will be explained in the sequel" or somehow explain them right away, otherwise we may lose the reader here.}

%\salwa{The above challenges and their prospective solution components are better highlighted in Fig.~\ref{fig:vision} where we illustrate the envisaged future \gls{ai}-aided \gls{6g-mn}s. This vision consists} of a high-performance terrestrial network segment that inherently supports operation and maintenance of smart harbors and industrial ports, and facilitates inland shipping and ubiquitous monitoring of near shore areas for improved human and marine life safety. Its non-terrestrial part encompasses a dynamic \gls{iab} infrastructure, which ensures connectivity bridges between vessels and the terrestrial segment. \salwa{As highlighted in Fig.~\ref{fig:vision}, an important component in this unified and all-in cellular solution for maritime networking is the powerful edge infrastructure strengthened with distributed learning and on-device intelligence.}

Pervasive system intelligence is vital for the evolution of maritime industry and its sustainable operation. In particular, real-time decision-making vastly improves port logistics and services. Through \gls{ai}-assisted remote control, an operator can digitally escort vessels safely to port. Smart fleet and asset tracking features can improve load distribution in ports, which decreases the volumes of carbon dioxide near the port areas. \gls{xr} applications for field engineers allow hands-on guidance from offsite support teams who can follow the operator's on-site view. All of these require lifelong learning where autonomous edge nodes (on-board the vessels or in dock areas) can participate in sensor data collection, processing, and sharing of resources for \gls{ml} model training. %~\cite{finland2020smartport}.
%Due to the intermittent nature of communication links in the integrated land, areal and sea segments, centralised cloud solution is not always feasible or rational from the perspective of latency tolerated by the intended applications and privacy of user data employed in learning.
%cohesion here?

Conventionally, \gls{dnn} algorithms are executed in the cloud where training data are preprocessed at the edge before being transferred to the cloud~\cite{hosseinalipour2020from}. The edge/fog computing infrastructures are intended to accommodate the needs of multiple \gls{dnn} models that require locality and persistent training. They also prevent the transmission of massive raw data over the network. Federated learning is a practical training mechanism wherein clients perform local \gls{ml} training and forward their results to an aggregator for further inference. Devices, edge nodes, and cloud servers can be equivalently deemed as clients. Under the risk of involving clients with poor channel conditions or limited energy supply, \gls{ml} model and client selection remains challenging in these distributed learning-based systems.

%Leveraging the proximity between edge nodes and parallel processing in on-demand \gls{iab} structures, the execution of tasks can be spatially and temporally distributed depending on the energy availability or cost-efficiency, while respecting application-specific latency and privacy constraints. With this approach, proximate vessels can offload unfinished tasks to close neighbors with better capabilities over sidelink or delay their computation to prevent interference growth.

%\subsection{Energy-Optimized Device-Level Solutions}\label{sec4.2}
\subsection{UE Capabilities in Device-Level Solutions}\label{sec4.2}

Mobile RedCap devices such as wearables can assume a central role in real-time monitoring and ubiquitous sensing of critical and highly dynamic processes in maritime environments. %including docking, loading and unloading of ships in ports, rescue missions on water, and virtually assisted or fully remote maintenance. Due to the dynamic environment they inevitably experience recurring radio link disruptions.
Unmanned aerial vehicles help create situational awareness for hinterland and smart fairway scenarios as well as provide remote technical support for container handling equipment. Enhanced with additional on-device learning and inference capabilities, these systems can utilize real-time data to offer deeper insights into energy-efficient maritime operations.
%For example, advanced port monitoring systems %with drone-enabled video monitoring and remotely controlled robots
%prevent port facility icing, keep ice loads from developing, and help in ice damage in Nordic regions, where ice accumulation inside ports causes difficulties for vessel maneuvering and slow down harbor operations. %~\cite{finland2020smartport}.
Beam misalignment in dynamic maritime environments may lead to significant data rate and energy efficiency degradation. \gls{dl}-based proactive beam management at the device side can help avoid this potential limitation. Whenever the line-of-sight link is not available, reinforcement learning allows the identification of optimal relay nodes in online fashion, even with limited prior knowledge of the environment. %~\cite{wang2020thirty}
However, selecting the optimal relay nodes is non-trivial in dynamic environments and when both in-band and out-of-band relaying options are available.

%Learning overhead is a crucial parameter to take into account when relying on \gls{ai}-aided frameworks. 
%The energy consumption of training \gls{ml} model depends on the global data set size and neural network architecture. 
%If a moving \gls{iab} node is installed on a floating platform or onboard of a ship, it can be permanently connected to a source of renewable energy, while drones and wearable devices would operate on battery. %However, the energy consumption of training \gls{ml} model depends on the global data set size and neural network architecture.
%Hence, device capabilities need to be considered when selecting suitable \gls{ml} models.

In systems relying on device-level solutions, UE capabilities are crucial when selecting suitable \gls{ml} models. For instance, if a moving \gls{iab} node is deployed on a floating platform or on-board a vessel, it can be connected to a source of renewable energy, while drones and wearable devices operate on battery. Therefore, not all devices in \gls{6g-mn}s are capable of training complex \gls{dnn} due to their limited storage, processing, or power capacity. 
Standalone compression techniques (such as pruning) have been optimized only for \gls{dnn} accuracy and without considering device energy consumption. 
By combining multiple compression techniques, one may derive compressed \gls{dl} models with desired trade-offs between performance and resource utilization~\cite{wang2020convergence}.
For instance, AdaDeep can automatically select various compression techniques to form a model according to not only device capability constraints but also application-driven requirements.

\subsection{Learning Delay and Network Reaction Time}\label{sec4.3}

%\olga{TODO: Distributed learning delay and DNN training time.}

In distributed learning under model or data split architecture, the involved nodes need to periodically communicate \gls{ml} model parameters over the network. The time to synchronize their results can grow significantly due to in-network transfer delays. This synchronization latency can become even higher when using \gls{ml} models such as \gls{dnn}s with thousands of parameters. Although several solutions were proposed to reduce the \gls{dnn} training times, the latter still depends on the used data samples and approximation functions. By properly selecting the nodes (e.g., depending on channel conditions and energy availability) and adjusting the \gls{ml} model parameters (e.g., learning rates and number of epochs in \gls{dnn}s), one can reduce the learning delay.

Learning delay and \gls{ml} convergence criteria are central in future \gls{ai}-aided \gls{6g-mn}s. %built with the principals of Open RAN alliance.
In learning-aided architectures, network reaction time can become a new key performance indicator that tells how soon a new system configuration can be enabled. It may be defined as the time between a parameter change (e.g., number of active users or link quality) and the network response time including \gls{ml} (re-)training and inference. %For instance, in the on-demand multi-hop topology formation challenge, this new metric will be determined with reference to the actual introduced changes in the multi-hop structure. 
Due to the limited communication ranges of high frequency radios, a potentially large number of hops may be required to connect two nodes of interest, which may cause an increase in the network reaction time under topology changes.

%Leveraging the proximity between edge nodes and parallel processing in on-demand \gls{iab} structures, the execution of tasks can be spatially and temporally distributed depending on the energy availability or cost-efficiency, while respecting application-specific latency and privacy constraints. With this approach, proximate vessels can offload unfinished tasks to close neighbors with better capabilities over sidelink or delay their computation to prevent interference growth.

\section{Conclusions and Outlook}\label{sec5}

The convergence of \gls{ai} and \gls{6g} allows to build sustainable \gls{ai}-aided networks for maritime communications. % with their unique requirements and challenges.
With the envisaged \gls{6g-mn}, the maritime industry can benefit from the enabling effects of digitalization and virtualization in reducing carbon dioxide emissions in ports and vessels. \gls{6g-mn}s permit the integration of terrestrial and non-terrestrial network segments, applications, and services in a holistic manner to accommodate the need for large-scale, sustainable, and on-demand system infrastructures. Due to network complexity and dynamics, \gls{ai}-aided solutions are indispensable for prompt and customized reactions of the network to demand fluctuations. In particular, deep learning techniques discussed in this article can tackle \gls{6g-mn} optimization challenges at different levels.
%These techniques can be used for maritime mesh topology management, traffic routing, and channel quality prediction.}
%While the standardization community focuses on defining interoperable building blocks in terms of capabilities, protocols, and procedures, distributed algorithms and additional incentives that engage end users and industries in further investigations can be addressed by the research community. Hence, the insights provided in this work can reinforce these incentives aiming to address the open questions and challenges in future intelligent \gls{6g-mn}s.

The challenges of efficient learning over \gls{6g-mn}s are shaped by the distinctive features of the rapidly changing maritime environment, remote operation with limited availability of energy and communication resources, and considerable learning delays in distributed systems. However, several approaches discussed in this work, such as reinforcement learning, can be further developed and employed to address these issues. The insights offered by this article motivate further research that can address the open questions and challenges in intelligent \gls{6g-mn}s.
%We anticipate that the answers to open questions related to different connectivity aspects in maritime networks can be obtained with the help of the proposed framework. In particular, deep and reinforcement learning techniques discussed above can tackle \gls{6g-mn} optimization challenges at different layers. These techniques can be used for maritime mesh topology design, traffic routing, mobility, and environment change prediction. While the standardization community focuses on defining interoperable building blocks in terms of capabilities, protocols, and procedures, distributed algorithms and additional incentives that engage end users and industries in further investigations can be addressed by the research community. Hence, the insights provided in this work can reinforce these incentives aiming to address the open questions and challenges in future intelligent \gls{6g-mn}s.

\vspace{-2 mm}

\section*{Acknowledgments}

The authors gratefully acknowledge funding from European Union's Horizon 2020
Research and Innovation programme under the Marie Sk{\l}odowska-Curie grant agreement No. 813278 (A-WEAR project). This work was also supported by the Academy of Finland (projects Emc2-ML, RADIANT, and IDEA-MILL).
G. Fodor was partially supported by the European Celtic project 6G-SKY with project ID C2021/1-9.

%\vspace{-2 mm}

\bibliographystyle{IEEEtran}
\bibliography{Main-arxiv}

%\newpage

\section*{Biographies}

\vspace{-10 mm}

\begin{IEEEbiographynophoto}{\sc{Salwa Saafi}}
(saafi@vut.cz) is a Ph.D. student at the Department of Telecommunications at Brno University of Technology, Czech Republic and the Unit of Electrical Engineering at Tampere University, Finland. She received her engineering degree (2017) in telecommunications from the Higher School of Communication of Tunis, Tunisia. Her research interests include cellular radio access technologies, future wireless architectures, and wearable applications.
\end{IEEEbiographynophoto}

%\vspace{-70 mm}

\begin{IEEEbiographynophoto}{\sc{Olga Vikhrova}}
(olga.vikhrova@tuni.fi) is a researcher at Tampere University, Finland. She received her Ph.D. (2021) in Information Engineering from University Mediterranea of Reggio Calabria, Italy, and M.Sc. (2014) in Information and Computer Science from Peoples' Friendship University of Russia (RUDN University), Russia. Her current research interests include distributed edge learning and computing, integrated access and backhaul networks, convergence of terrestrial and non-terrestrial networks.
\end{IEEEbiographynophoto}

%\vspace{-70 mm}

\begin{IEEEbiographynophoto}{\sc{Gabor Fodor}}
(gaborf@kth.se) received his Ph.D. degree in electrical engineering from the Budapest University of Technology and Economics in 1998, his Docent degree from KTH Royal Institute of Technology, Stockholm, Sweden, in 2019, and his D.Sc. degree from the Hungarian Academy of Sciences in 2019. He is currently a master researcher with Ericsson Research and an adjunct professor with KTH Royal Institute of Technology. He was a co-recipient of the IEEE Communications Society Stephen O. Rice Prize in 2018. He is serving as the Chair of the IEEE Communications Society Full Duplex Emerging Technologies Initiative and as an Editor for IEEE Transactions on Wireless Communications and IEEE Wireless Communications.
\end{IEEEbiographynophoto}

%\vspace{-70 mm}

\begin{IEEEbiographynophoto}{\sc{Jiri Hosek}}
(hosek@vut.cz) received his M.S. and Ph.D. degrees in electrical engineering from the Faculty of Electrical Engineering and Communication at Brno University of Technology (BUT), Czech Republic, in 2007 and 2011, respectively. He is currently an Associate Professor (2016) and Deputy Vice-Head for R\&D and International Relations at the Department of Telecommunications (2018), BUT. Jiri is also coordinating the WISLAB research group (2012), where he deals mostly with industry-oriented projects in the area of IoT and home automation services. Jiri (co-)authored more than 130 research works on networking technologies, wireless communications, quality of service, quality of experience, and IoT applications.
\end{IEEEbiographynophoto}

%\vspace{-70 mm}

\begin{IEEEbiographynophoto}{\sc{Sergey Andreev}}
(sergey.andreev@tuni.fi) is an associate professor of communications engineering and Academy Research Fellow at Tampere University, Finland. He has been a Visiting Senior Research Fellow with King's College London, UK (2018-20) and a Visiting Postdoc with University of California, Los Angeles, US (2016-17). He received his Ph.D. (2012) from TUT as well as his Specialist (2006), Cand.Sc. (2009), and Dr.Habil. (2019) degrees from SUAI. He served as lead series editor of the IoT Series (2018-21) for IEEE Communications Magazine and as editor for IEEE Wireless Communications Letters (2017-19).
\end{IEEEbiographynophoto}

\end{document}